\documentclass{iau}
\pdfoutput=1
\usepackage{graphicx}
\usepackage{psfig}
\usepackage{epsf}
\usepackage{graphicx} 

\title[Early mass assembly of the Milky Way halo] 
{The early gaseous and stellar mass assembly of Milky Way-type galaxy halos}

\author[Gerhard Hensler \& Mykola Petrov]  
{Gerhard Hensler$^{1,2}$  
\and
Mykola Petrov$^1$
}

\affiliation{$^1$Dept. of Astrophysics, Univ. of Vienna,
Tuerkenschanzstr. 17, 1180 Vienna, Austria \\
email: gerhard.hensler@univie.ac.at \\
$^2$Nat. Astron. Obs. of Japan, 2-21-1 Osawa, Mitaka-shi, 
Tokyo 181-8588, Japan}

\pubyear{2016}
\volume{317}  
\pagerange{119--126}
\date{?? and in revised form ??}
\jname{The General Assembly of Galaxy Halos: 
Structure, Origin and Evolution }
\editors{eds. A. Bragaglia \& M. Arnaboldi}

\begin{document}

\def\HI{H{\sc i} }
\def\HII{H{\sc ii} }
\def\Ha{{\rm H}\alpha }
\def\rSF{\rho_{\rm SF} }
\def\rg{\rho_{\rm g} }
\def\Msun{M_{\odot} }
\def\sfr{$\Msun yr^{-1}$}
\def\Mpc2{\Msun/pc^2}
\def\tff{\tau_{\rm ff} }
\def\tSF{\tau_{\rm SF} } 
\def\e{\epsilon} 
\def\eSN{\epsilon_{\rm SN} }
\def\eSF{\epsilon_{\rm SF} }
\def\OH{{12\-+log(O/H)} }
\def\lNO{log(N/O) }
\def\cd{chemo-dynamical }
\def\aFe{{$\alpha$/Fe} }
\def\LCDM{$\Lambda$CDM}
\def\aap{\textit{Astron.Astroph.}}
\def\apj{\textit{Astroph.J.}}
\def\aj{\textit{Astron.J.}}
\def\nat{\textit{Nature}}
\def\mn{\textit{MNRAS}}
\def\araa{\textit{Ann. Rev. A{\rm \&}A}}

\maketitle

\begin{abstract}
How the Milky Way has accumulated its mass over the Hubble
time, whether significant amounts of gas and stars were accreted
from satellite galaxies, or whether the Milky Way has experienced
an initial gas assembly and then evolved more-or-less in isolation
is one of the burning questions in modern astronomy, because it
has consequences for our understanding of galaxy formation
in the cosmological context. Here we present the evolutionary
model of a Milky Way-type satellite system zoomed into a
cosmological large-scale simulation. Embedded into Dark Matter 
halos and allowing for baryonic processes these \cd
simulations aim at studying the gas and stellar loss from the 
satellites to feed the Milky Way halo and the stellar chemical 
abundances in the halo and the satellite galaxies.  
\end{abstract}

\firstsection 

\section{Introduction}

In the past half century, by means of more sensitive observations of 
the Milky Way (MWG) stellar halo its formation was interpreted by two 
different processes: 
The first prefers the monolithic collapse \cite{egg62}, and the 
second is the accretion model by \cite{sea78}. At the same time, 
\cite{whi78} proposed their Cold Dark Matter (CDM) hierarchical 
clustering paradigm in which galaxies are results from cooling and fragmentation of residual gas within the transient potential wells 
provided by the DM. In this framework galaxy formation proceeds in 
a "bottom up" manner starting with the formation of small clumps 
of gas inside DM subhalos, which then merge hierarchically into 
larger systems (\cite{blu84,spr05}).

CDM simulations of cosmological structure and galaxy formation 
predict the existence of a large number of such DM subhalos 
surrounding massive DM gravitational potentials.
These subhalos should serve as the DM progenitors of dwarf galaxies 
(DGs) which indeed permeate the Local Group (LG), most of them 
concentrated as satellites around the MWG and M31. The closer to a
mature galaxy they live, the more gas-free they are like elliptical 
DGs and are, therefore, called dwarf spheroidals (dSphs).
Because of their low surface brightness, though even close to the MWG, 
for a long time only a few of them could be observed in the range 
of M$_V$ = -14$^m$ to -10$^m$ separated clearly from Globular Clusters.  
Their number increased over the last years thanks to systematic 
surveys like SDSS shifting the lower brightness limit by the recently 
discovered ultra-faint DGs (UFDs) to almost -2$^m$ (see e.g. 
\cite{bel10,bel14}) so that they extend the DG sequence to its 
faintest end. 

Theoretically already expected and verified by numerical simulations
(\cite{joh08}), satellites in the neighbourhood and, thereby, in the 
tidal field of mature galaxies lose continuously gas and stars, the
later observable as tidal streams (\cite{lyn95,jer13}). 
Due to their loss of orbital energy and angular momentum their fate 
is the partly disruption and their death as individuals is the 
accretion by their mature galaxy.
This scenario of tidal disruption is at present most strikingly 
demonstrated by the Sagittarius DG (\cite{iba94}) with its tidal tails 
wrapped around the MWG (\cite{maj03}). 

The CDM merging hypothesis requires the infall and accumulation of 
the MW mass not only by DM subhalos and gas but also by stars. 
If this ''mining'' of the halo with dSph stars (\cite{sal08}) has 
happened in the very early epoch with the first stars only, no 
differences at the low Z end will tell us about as long as the MW 
halo stars are formed from the same gaseous substrate. 
At larger metallicities the \aFe ratio of dSphs, however, declines 
already (due to supernova type Ia (SNIa) enrichment) while the halo 
stars are systematically at the constant value of SNII enrichment at 
the same [Fe/H] (\cite{she03,tol03,ven04,koc08};
see also reviews by \cite{koc09} and \cite{tol09}). 
This fact allows to pin-down that progenitors of present-day dSphs 
are not the expected building blocks of the galactic halo and to
explain the lack of the observed number of stellar streams. 
From kinematics of halo stars, however, a dichotomy is found
by \cite{car07} and \cite{bell08}, one regular population in the inner 
region of about 10 kpc radius and an outermost heterogeneous and decoupled 
halo population most plausibly accreted from disrupted satellites.
The metallicity distribution function (MDF) of the UFDs suggests that 
these tiny systems contain a larger fraction of extremely metal-poor 
stars than the MW halo does (\cite{kir08}) and witness the chemical
imprint of the interstellar medium (ISM) when the Universe was 
less than 1~Gyr old.

Detections of hyper metal-poor stars (\cite{bee05}) in the galactic halo 
and their peculiar element abundances (see e.g. \cite{fre05}) 
opened a new field of galactic archeology, namely, modelling the element 
production by the first stars in the halo as well as in the UFDs 
towards understanding the zero metallicity nucleosynthesis 
and studying the formation of the halo. Chemical evidences for this 
scenario, especially in the metal-poor stellar content of the galactic
halo, is mentioned also by \cite{fre10}, \cite{fre12}, and others. 
On the other hand, it is currently unclear how the metal-poor MDF tail
of the classical DGs, in which extremely metal-poor stars are
absent, compares with that of the halo and the UFDs.

\section{Modeling the Milky Way satellite system}

In contrast to the DM evolution of subhalos treated by pure N-body 
simulations, the evolution of the baryonic component is much more 
complex because of the physical processes at work, such as star 
formation (SF), gas cooling, dissipation, energy and mass feedback. 
Baryonic gas loses kinetic energy 
dissipatively and thermal energy by radiation leading to cooling 
and gravitational collapse, while stellar radiation and winds 
as well as SNe lead to energy and chemical feedback.
Almost all modelling up to now is dedicated to investigate the
effects of various processes on the dSph evolution separately
or the dSph evolution as an isolated system.

Although the gasdynamical simulations of dSphs advanced from 1D \cd\ 
models by \cite{hen04} to 3D (see e.g. Smooth-Particle Hydrodynamics (SPH)
models by \cite{rev09,pas11}), they mainly lack not only of a self-consistent treatment of various internal processes, but focussed 
on particular aspects only. Their results do not deviate too much from 
observational data, however, the system of satellites is exposed to
a whole bunch of external processes also, like e.g. ram-pressure 
(\cite{may07}) and tidal (\cite{rea06a}) stripping, gas accretion, 
and further more. If these are not taken into account but simulations 
are limited to isolation DGs the models cannot allow a reliable 
trace-back of the evolution of any dSph galaxy.

\begin{figure}[!h]
\begin{center}
  \includegraphics[width=10cm, height=7.5cm]{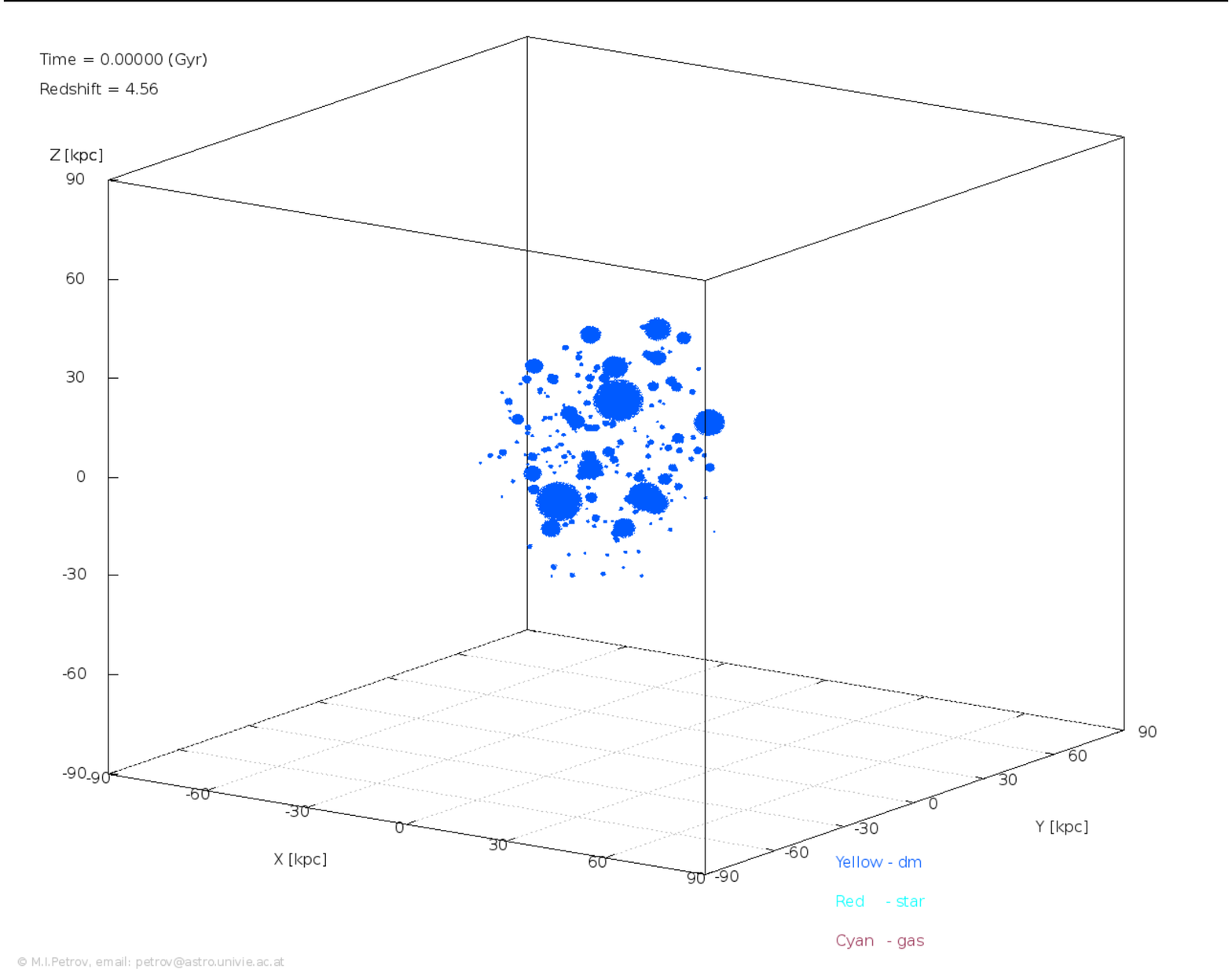}  
  \includegraphics[width=10cm, height=15cm]{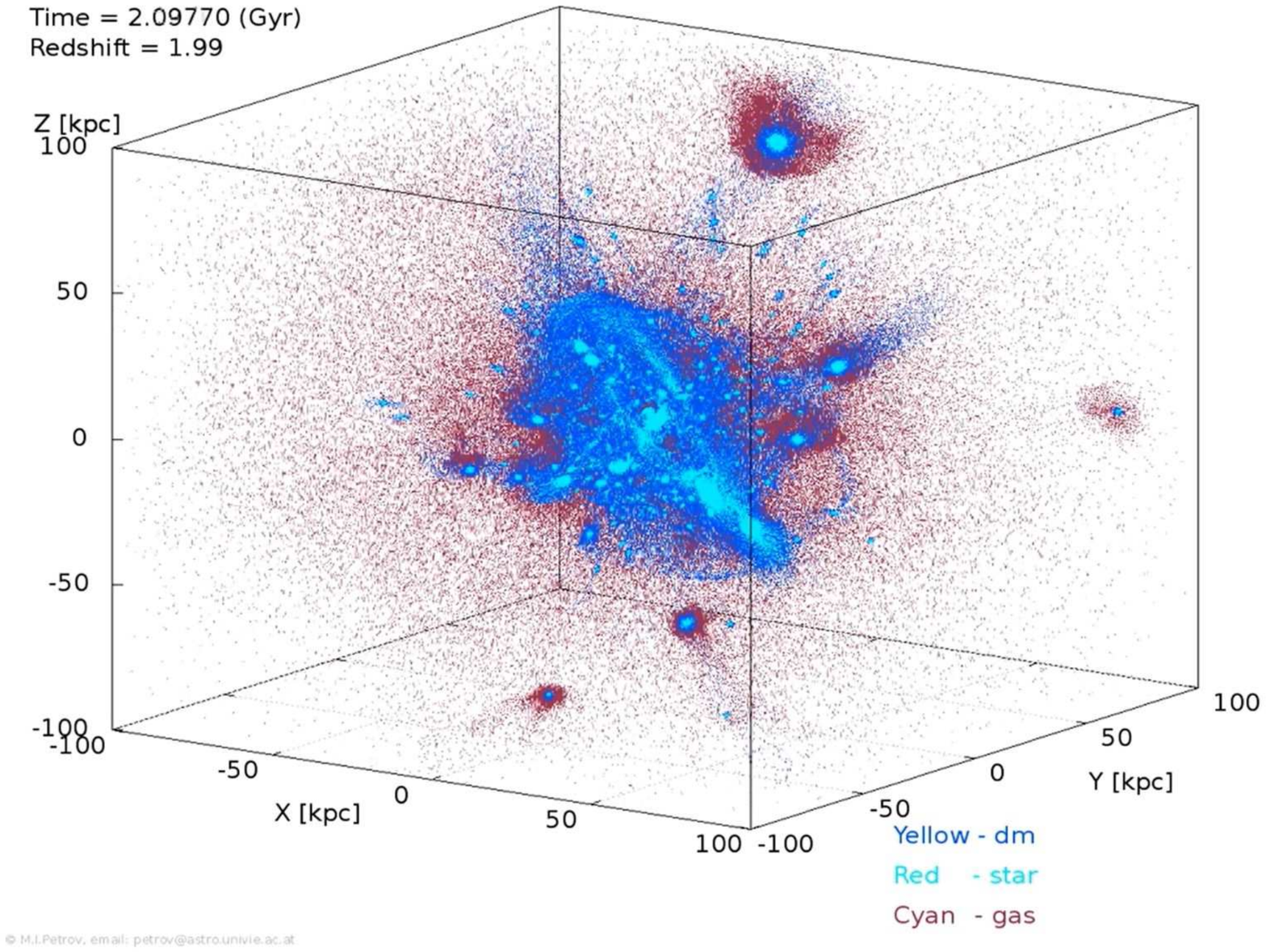} 
\end{center}
\vspace{-7cm}  
\caption{
Cubes of 200 kpc length around the Milky Way (at their center). 
{\it upper panel:} 
Initial conditions of the Milky Way's satellite system:    
Distribution of Dark Matter (DM) subhalos within a sphere 
of 40 kpc radius around the Milky Way at redshift z=4.56. 
{\it lower panel:} 
Snapshot of the satellites' dynamical evolution 
2.1 Gyr after the numerical onset, i.e. at redshift z=2. 
The DM subhalos are filled with 17\% baryonic gas mass, 
form stars, and lose mass of all constituents due to 
various effects (see text). 
}
   \label{fig:sat}
\end{figure}

In addition, simple non-dynamical considerations as performed to 
understand the chemical evolution 
(\cite{lan06,lan10,pra08,kir11a,kir11b}) 
provide only a limited understanding of the real evolution of dSphs.

\cite{fon06} investigated the nature of the progenitors of the stellar 
halo for a set of MWG-type galaxies and studied the chemical enrichment
patterns in the context of the CDM model with a combination of
semi-analytic prescriptions. They concluded that the difference in chemical 
abundance patterns in local halo stars versus surviving satellites arises 
naturally from the predictions of the hierarchical structure formation 
in a CDM universe.

Here we present a model of the early evolution and mass assembly
history of the MWG's halo by the system of satellite galaxies
treated in the gravitational field of the MWG. For this purpose
we select a MWG-like DM host halo from the cosmological $\Lambda$CDM 
simulation Via Lactea II (\cite{die08}). The constraints are, that it 
does not undergo a major merger event over the Hubble time and that
sufficient subhalos exist which allow the accretion by the host galaxy. 
For the simulations an advanced version of the single-gas 
\cd SPH/N-body code is applied, treating the production 
and chemical evolution of 11 elements.

Since the acceptable computational time limits the number of gas 
particles to two million and the DM particles to the same order
and because we aim at reaching a mass resolution of 10$^3 \Msun$ 
per SPH particle, only 250 subhalos in the total-mass range 
of $10^6 < M_{sat}/\Msun < 6\times 10^8$ could be followed from 
redshift $z=4.56$ with its baryonic content.
Unfortunately, this fact limits the radius of consideration to 
40 kpc around the MWG's center of mass.
In order to study the construction of the MWG halo by accretion
of subhalos with both gas and stars, as a first step, the \cd 
evolution of the dSph system is followed for the first 2 Gyr, 
i.e. until redshift $z=2$ (see Fig.\ref{fig:sat}).

Starting with a $10^4$ K warm gas of 17\% of the subhalo masses
in virial equilibrium and under the assumption that 
re-ionization is improbable to have affected the LG dSphs 
(\cite{gre04}), cooling allows the gas particles to achieve 
SF conditions in all satellites, but its efficiency directly 
depends on the mass of a satellite and its dynamical history 
(e.g. merging with other satellites or disruption by the MWG 
gravitational potential).

\section{The Milky Way halo formation}
 
\begin{figure}[!h]
\begin{center}
\includegraphics[width=10cm, height=15cm]{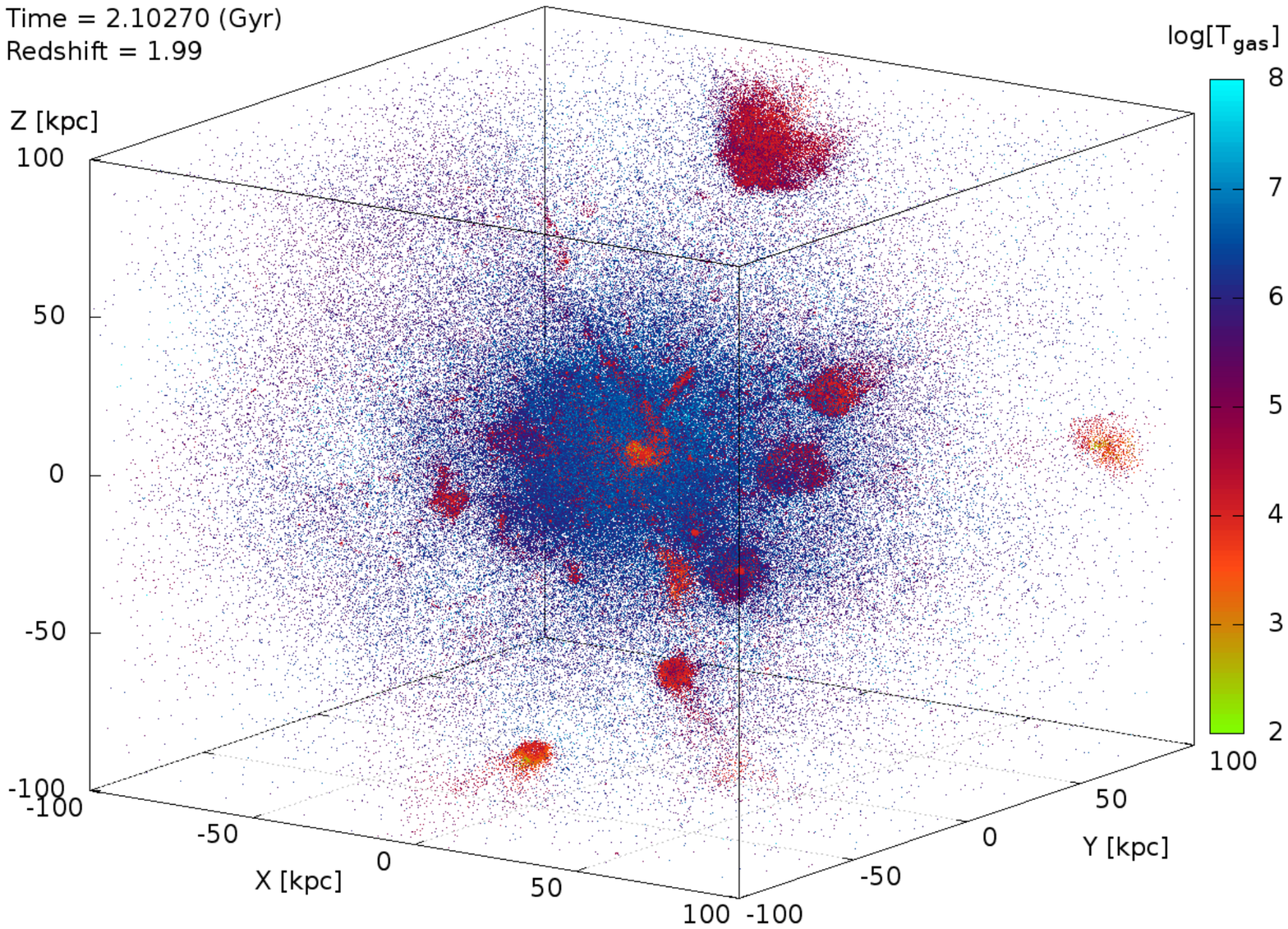} 
\end{center}
\vspace{-7cm}
\caption{Snapshot of the gas distribution 2.1 Gyr after the numerical
onset, i.e. at redshift z=2.76. The gas temperature is coloured according
to the right colour panel.
}
   \label{fig3}
\end{figure}

Fig. 1 shows the evolution of the DM-gas-stars composit of the 
satellite galaxies. The SF starts in all satellites almost 
simultaneously, then ceases for the lowest mass objects, while it 
continues in more massive ones with fluctuations due to gas loss
but also interactions with other objects. dSphs develop their 
stellar components and element abundances dependent on the distance 
from the MWG.
Gas is pushed out from low-mass dSphs by their internal stellar energy
release and lost from massive dSphs more by the tidal force. 
Both effects feed the MWG gas halo by pre-processed hot gas (fig. 2).
Inherently as an additional effect, dSphs also get rid of their gas 
by their motion within the bath of their lost hot gas. Stars are also 
disrupted from the satellites and accumulate in the MWG halo at the early stages from all objects with low metallicity. Lateron, only stars
from the massive satellites contribute due the cessation of SF in the
less-massive systems.  
In total, 1.88$\times 10^8 \Msun$ of gas and 9.53$\times 10^7 \Msun$ of stellar mass are torn off from the satellite galaxies and got bound 
to the Galactic halo. From the same demolishing effect additionally
2.63$\times 10^9 \Msun$ of DM mass fed the MWG.

\begin{figure}[!h]
\begin{center}
\includegraphics[width=11cm]{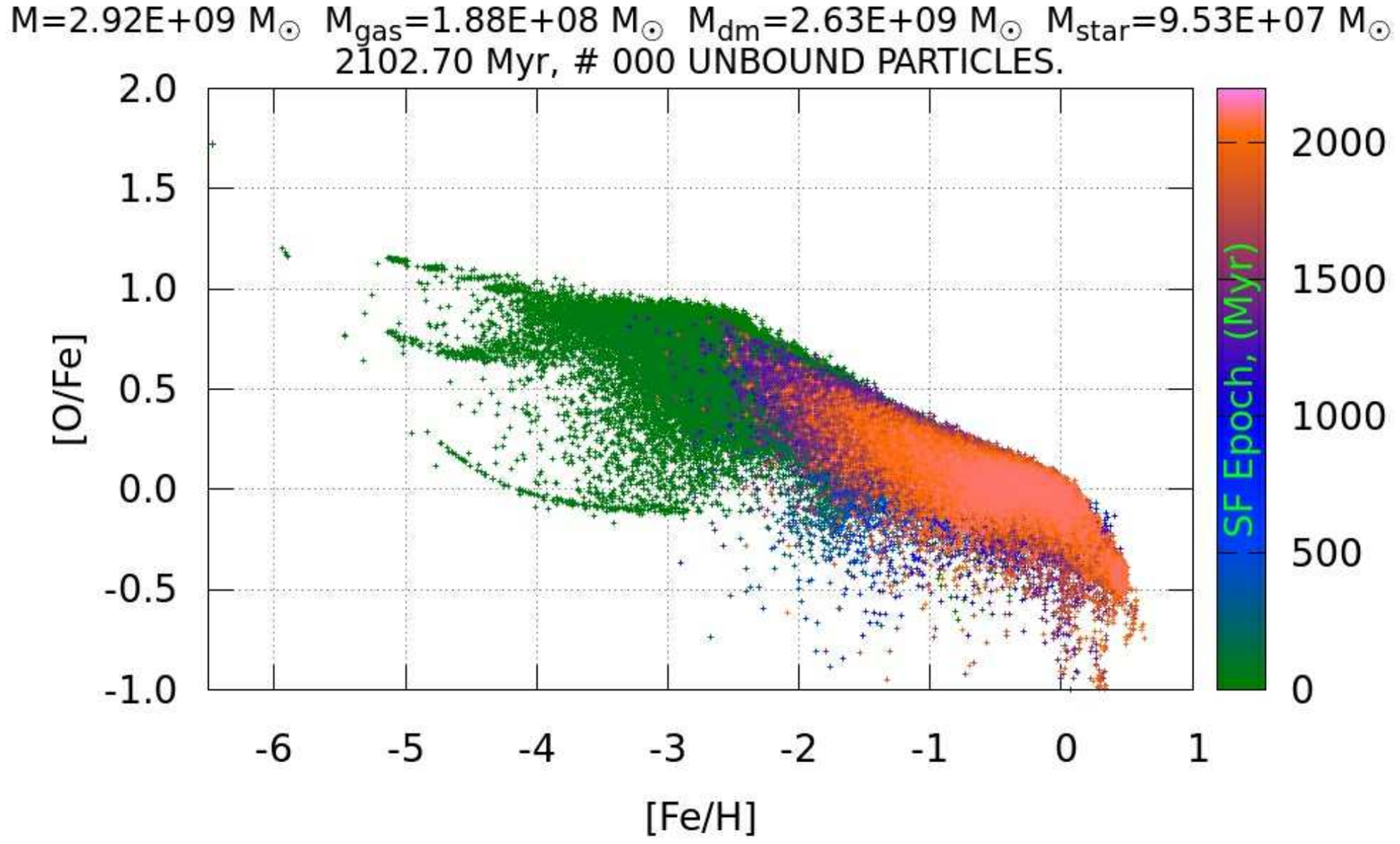} 
\end{center}
\vspace{-8cm}
\caption{
[O/Fe] vs. [Fe/H] distribution of star and gas that became unbound 
from the satellite galaxies within the first 2.1 Gyr after the
model onset. The values of the element ratios are coloured according
to the time when the stars and gas are dissolved from the satellites.
The emerging stripes are artefacts by used values from tables
during the first enrichment episode. 
}
   \label{fig4}
\end{figure}

For the first 0.1 Gyrs of the simulation there is a considerable
variance of stellar oxygen abundance in the whole system 
$(-5. \leq [O/H] \leq -0.5)$ reflecting the very inhomogeneous 
production and distribution of enriched gas.
After 0.1 Gyrs merging of the satellites' ISM promotes the mixing 
of heavy elements. Finally, almost complete recycling of the gas 
erases the abundance inhomogeneities so that oxygen in stars 
converges to $-1. \leq [Fe/H] \leq 0.$ with a small dispersion 
(fig. 3). These high abundances show that the too efficient 
metal-enrichment in a single-gas phase treatment has to be relaxed 
by a more realistic \cd multi-phase prescription of the ISM 
(\cite{liu15}).

\begin{acknowledgments}
The authors are grateful to Simone Recchi for his contributions and continuous discussions on the chemical enrichment.
This work was partly supported by the Austrian Science Foundation 
FWF under project no. P21097.
\end{acknowledgments}

\end{document}